\documentclass{sig-alternate-05-2015}

\usepackage[
  pass,
]{geometry}

\newfont{\mycrnotice}{ptmr8t at 7pt}
\newfont{\myconfname}{ptmri8t at 7pt}
\let\crnotice\mycrnotice%
\let\confname\myconfname%

\usepackage{url}
\usepackage[neverdecrease]{paralist}
\usepackage[caption=false]{subfig}
\usepackage{graphicx}
\usepackage{booktabs}
\usepackage{graphics}
\usepackage{tabularx}
\usepackage{xspace}
\usepackage{graphicx}
\usepackage{booktabs}
\usepackage{color}

\usepackage{amsfonts}
\usepackage{mdwlist}
\usepackage{multirow}
\usepackage{amsmath}
\usepackage{enumitem}
\usepackage{calc}
\usepackage[ruled,vlined]{algorithm2e}

\usepackage{amssymb}
\usepackage{pifont}

\makeatletter
\def\BState{\State\hskip-\ALG@thistlm}
\makeatother

\newcommand{\urlwofont}[1]{ \urlstyle{same}\url{#1} }

\begin{document}

\permission{Permission to make digital or hard copies of part or all of this work for
personal or classroom use is granted without fee provided that copies are
not made or distributed for profit or commercial advantage and that copies
bear this notice and the full citation on the first page. Copyrights for
third-party components of this work must be honored. For all other uses,
contact the owner/author(s).}
\conferenceinfo{Neu-IR '16 SIGIR Workshop on Neural Information Retrieval}{July 21, 2016, Pisa, Italy \\
{\mycrnotice{\copyright 2016 Copyright held by the owner/author(s).}}}

\title{Representing Documents and Queries as Sets of Word Embedded Vectors for Information Retrieval}

\numberofauthors{2} 
\author{
\alignauthor
Dwaipayan Roy \\
       \affaddr{CVPR Unit}\\
	   \affaddr{Indian Statistical Institute} \\
       \affaddr{Kolkata, India}\\
       \email{dwaipayan.roy@gmail.com}
\and
\alignauthor
Debasis Ganguly\\
       \affaddr{ADAPT Centre, School of Computing} \\
       \affaddr{Dublin City University} \\
       \affaddr{Dublin, Ireland}\\
       \email{dganguly@computing.dcu.ie}
\and 
\alignauthor
Mandar Mitra\\
       \affaddr{CVPR Unit}\\
	   \affaddr{Indian Statistical Institute} \\
       \affaddr{Kolkata, India}\\
       \email{mandar@isical.ac.in}
\alignauthor
Gareth J.F. Jones\\
       \affaddr{ADAPT Centre, School of Computing}\\
       \affaddr{Dublin City University}\\
        \affaddr{Dublin, Ireland}\\
       \email{gjones@computing.dcu.ie}
}

\maketitle

\begin{abstract}
A major difficulty in applying word vector embeddings in information retrieval is in devising an effective and efficient strategy for obtaining representations of compound units of text, such as whole documents, (in comparison to the atomic words), for the purpose of indexing and scoring documents. Instead of striving for a suitable method
to obtain a single vector representation of a large document of text, we 
aim to develop a similarity metric that makes use of the similarities between the individual embedded word vectors in a document and a query. More specifically, we represent a document and a query as sets of word vectors, and
use a standard notion of similarity measure between these sets, computed as a function of the similarities between each constituent word pair from these sets. We then make use of this similarity measure in combination with standard information retrieval based similarities for document ranking. The results of our initial experimental investigations show
that our proposed method improves MAP by up to $5.77\%$, in comparison to standard text-based language model similarity, on the TREC 6, 7, 8 and Robust ad-hoc test collections.

\end{abstract}




\terms{Theory, Experimentation}


\section{Introduction} 
\label{sec:intro}

Embedding words as real-valued vectors enables the application of neural network (NN) based approaches for several supervised natural language processing (NLP) tasks, such as sentiment analysis, named entity recognition, semantic role labelling etc. \cite{Collobert:2011}. The real-valued vectors are fed into the input layers of NNs designed for various supervised
tasks. The word embeddings themselves are obtained by unsupervised pre-training from a large corpus of text,
by making use of recurrent neural networks (RNNs) \cite{Bengio:2003, Mikolov13}. The specific technique of
word embeddings, known as \emph{word2vec}, uses the core idea of \emph{negative sampling}
which massively increases the efficiency of the embedding process so as to make it scalable for
large document collections \cite{Mikolov13}. The main idea behind \emph{negative sampling} is to design an objective function that seeks to maximize the similarity of the vector representation of the current word with
another word sampled from the context of it, i.e. from a word window of a preset length around the current word. The objective function also seeks to minimize the similarity of the vector representation of the current word with another word drawn randomly from outside its context; hence the name \emph{negative sampling} \cite{Mikolov13}.

A useful feature of word vectors pre-trained 
using word2vec is that vector addition (component wise addition) of two or more word vectors results in a vector that is close to the semantic meaning of the constituent words.
This property of the embeddings has been used to address the proportional analogy task in NLP, such
as predicting that `Berlin' is the capital of `Germany' given that `Paris' is the capital of `France', since
$vec(Berlin)$ is close to the vector $vec(Paris)+vec(France)-vec(Germany)$.
Word embedding thus encapsulates the useful information about the context of a word and effectively
represents the semantic information of a word. It is thus tempting to use the real-valued vector representations
of words to
represent documents and queries in information retrieval (IR) IR and defining similarity measures between them. 

However, a major difficulty in applying word vector embeddings in IR is to be able to devise an effective strategy for
obtaining representations of compound units of text (in comparison to the atomic words), such
as passages and documents for the purpose of indexing.
Adding word vectors for deriving composed meanings has been shown to be beneficial for small units of text, e.g.
for predicting query intent in search sessions \cite{Mitra:2015}. 
However, this approach of adding the constituent word vectors
of a document to
obtain the vector representation of the whole document is not likely to be useful, because
the notion of compositionality of the word vectors works well when applied over a relatively small number of words.
A vector addition for composition does not scale well for a larger unit of text, such as passages or full documents, because of the broad context present within a whole document. A possible solution is to use an extension of `word2vec', commonly called `doc2vec', to
obtain distributed representations of arbitrary large units of text, such as paragraphs or documents, where in addition to
learning the vector representations of words, the RNN also learns of the vector representation of the current document with a modified objective function \cite{LeM14}. Unfortunately, training `doc2vec' on a collection
of a large number of documents (as often encountered in IR) is practically intractable both in terms of speed and memory resources.

We address this issue from a different perspective. Instead of striving for a suitable method for obtaining a single vector representation of a large document of text and a query, so as to use the inner product similarity between them for scoring documents, we seek
to 
develop a suitable similarity (or inverse distance) metric which makes use of the each individual embedded word vector in a document
and a query. More specifically, we represent each document (and a query) as a set of word vectors, and use a standard notion of similarity measure between these sets.
Typically, such set based distance measures, such as the single-link, complete-link, average-link similarities etc. are used in hierarchical agglomerative clustering (HAC) algorithms \cite{manning2009}. Generally speaking, these similarities are computed as functions of the individual similarities between each constituent word vector pair of a document and a query set. We then combine this word-vector based similarity measure with a standard text based IR similarity measure in order to rank the retrieved documents in response to a query.

The remainder of the paper is organized as follows.
In Section \ref{sec:representation}, we discuss how documents
and queries are represented as sets of word vectors so as to
compute the similarities between them. Section \ref{sec:indexing}
describes the index construction procedure to store and make use
of the additional word vectors. Section \ref{sec:eval} evaluates
our proposed approach of document ranking with the word vector
based similarity. Finally, Section \ref{sec:concl} concludes the
paper with directions for future work.


\section{Document Representation} 
\label{sec:representation}


\subsection{Documents as Mixture Distributions}

The `bag-of-words' (BoW) representation of a document within a collection treats the document as a sparse vector in the term space
comprised of all terms in the collection vocabulary.
The vector representation of each word, obtained with a word embedding
approach, makes provision to view a document as a `bag-of-vectors' (BoV)
rather than as a BoW.
With this viewpoint, one may imagine that a document is
a set of words with one or more clusters of words, each cluster
broadly representing a topic of the document.
To model this behaviour, it can be assumed that a document is
a probability density function that generates the observed
document terms.

In particular, it is convenient to represent
this density function as a mixture of Gaussians of $p$ dimensions so that each word vector (of $p$ dimensions) is an observed sample drawn from this mixture distribution. With the observed document words,
this probability can be estimated with an expectation maximization (EM) algorithm, e.g. K-means clustering of the observed document vectors.
The observed query vectors during retrieval can be considered
to be points that are also sampled from this mixture density
representation of a document. Documents can then be ranked by
the posterior likelihood values of generating query vectors from their
mixture density representations.

More specifically, let the BoW representation of a document $d$ be $W_d = \{w_i\}_{i=1}^{|d|}$, where $|d|$ is the number of unique words in $d$ and $w_i$ is the $i^{th}$ word. The BoV representation of $d$ is the set $V_d = \{x_i\}_{i=1}^{|d|}$, where $x_i \in \mathbb{R}^p$ is the vector representation of the word $w_i$.
Let each vector representation $x_i$ be associated with a latent
variable $z_i$, which denotes the topic or concept of a term and is an integer between $1$ and $K$, where $K$, being a parameter, is the total number of topics or the number of Gaussians in the mixture distribution.
These latent variables, $z_i$s, can be estimated by an EM based
clustering algorithm such as K-means, where after the convergence of K-means on the set $V_d$, each $z_i$ represents the cluster id of each
constituent vector $x_i$. Let the points $C_d = \{\mu_k\}_{k=1}^{K}$ represent the $K$ cluster centres as obtained by the K-means algorithm.

\subsection{Query Likelihoods}

The posterior likelihood of the query to be sampled from the
$K$ mixture model of Gaussians, centred around the $\mu_k$ centroids,
can then be estimated by the average distance of the observed query points, $q=\{q_i\}_{i=1}^{|q|}$, from the centroids of the clusters.
\begin{equation}
sim(q,d) = \frac{1}{K|q|} \sum_{i}\sum_{k}q_i\cdot \mu_k  \label{eq:avgsim}
\end{equation}
In Equation \ref{eq:avgsim}, $q_i\cdot \mu_k$ denotes the inner product
between the query word vector $q_i$ and the $k^{th}$ centroid vector
$\mu_k$. This measure is commonly known as the \emph{centroid similarity} or \emph{average inter-similarity} in the literature, where it finds application to measure how similar two sets are during hierarchical agglomerative clustering (HAC) \cite{manning2009}.

Although there are multiple set-based similarity measures, such as
the single-link, complete-link similarities etc., we report the average-link
similarity in our experiments because it produced the best results. A likely
reason for this metric to produce the best results is that it is not largely affected by the presence of outliers like single-link or complete-link
algorithms (for more details, see the discussion in Chapter 17 of the book \cite{manning2009}).

Intuitively speaking, the notion of set-based similarity is able to make use
of the semantic distances between the constituent terms
of a document and a given query so as to improve the rankings
of retrieved documents. Note that a standard text-based
retrieval model can only make use of the term overlap statistics between documents and queries, and cannot utilize the semantic
distances between constituent terms for scoring documents.

\begin{figure}[t]
    \subfloat[] {\label{fig:case1_a}
    \includegraphics[width=0.49\columnwidth]{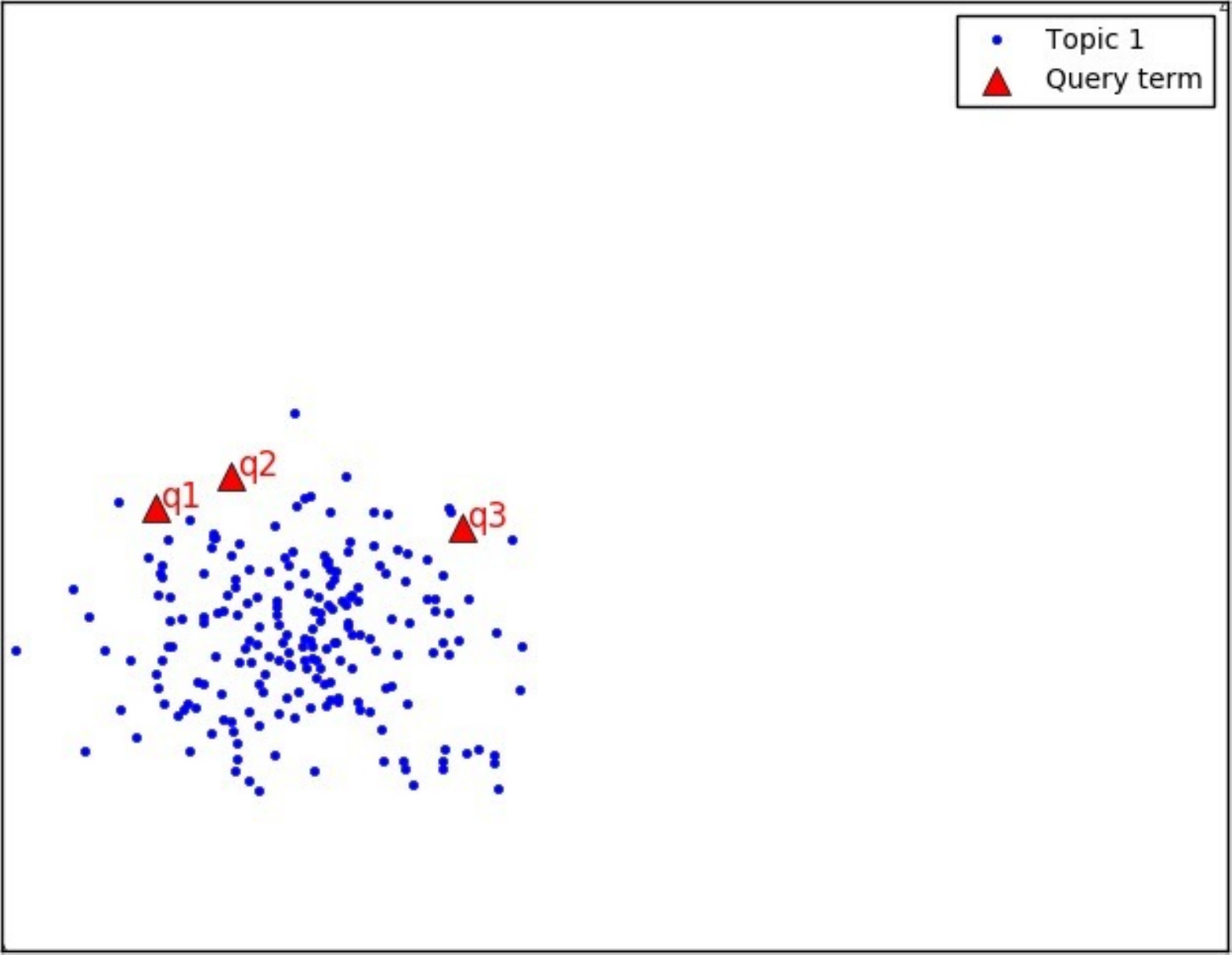}}
    \subfloat[] {\label{fig:case1_b}\includegraphics[width=0.49\columnwidth]{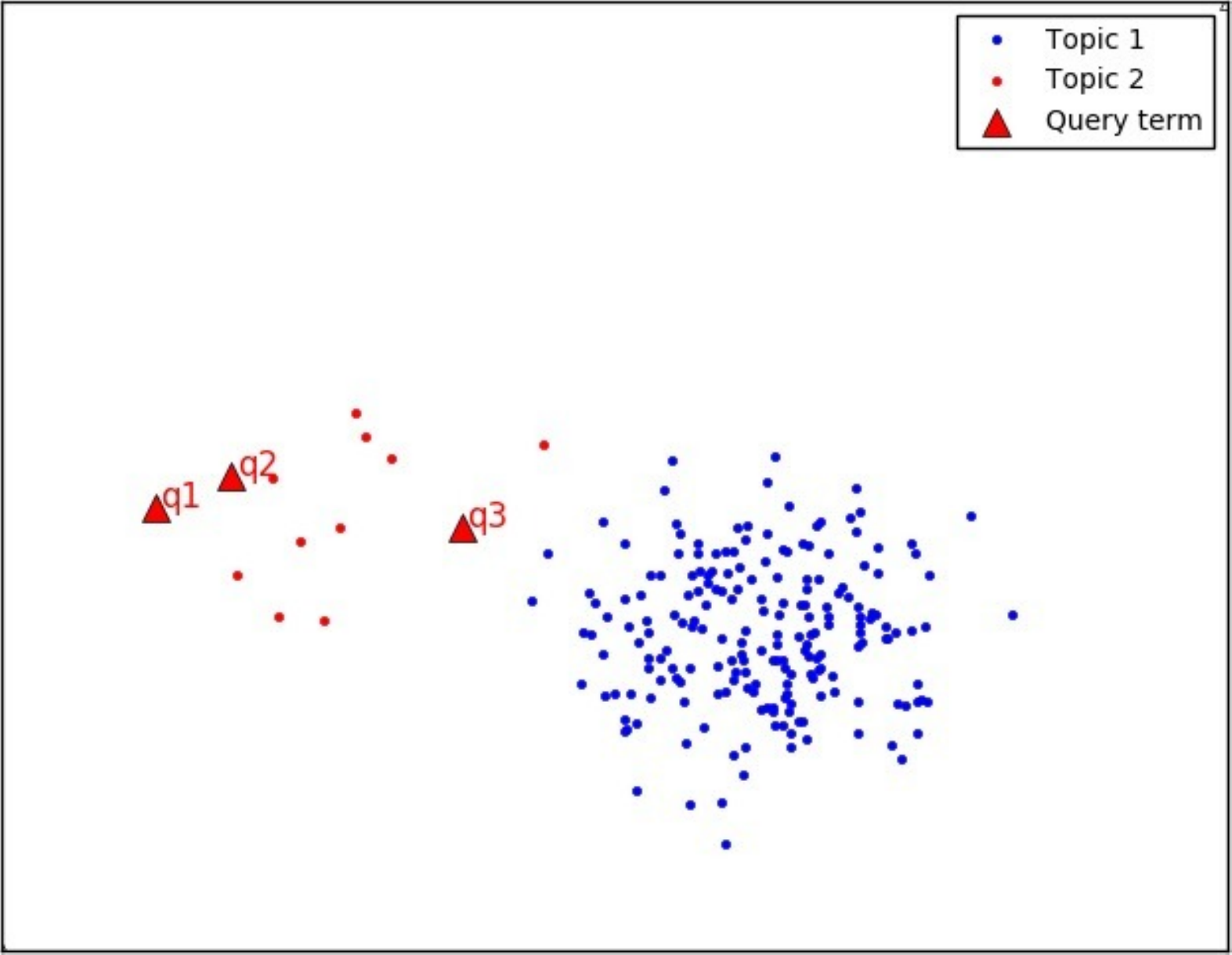}}    
    {\caption{Two example scenarios of (predominantly) single-topical documents, where the document on the left has a higher similarity to the query than the document on the right.}\label{fig:case1}}
\end{figure}

\subsection{Illustrative Examples}

We now demonstrate the key idea of the usefulness of the set-based similarity of the constituent word vectors of documents and queries with illustrative examples.
Consider the documents of Figure \ref{fig:case1}, where
for illustrative purposes, we assume that the $p=2$, i.e., each word is embedded in a two dimensional space. The individual word vectors of a document are shown with
blue dots, whereas the query points are shown with red triangles. Note that the document in Figure \ref{fig:case1_a} has one cluster
and all the three query points are relatively close to
the centroid of this cluster. In contrast, the document
in Figure \ref{fig:case1_b} is comprised of two clusters (one blue and the other red). In this case, the query terms are relatively far
away from the central theme of the document, i.e. the
position of the centroid vector of the predominant topic
of the document, i.e the centroid of the blue words.
This indicates that the posterior query
likelihood for the document in Figure \ref{fig:case1_b}
is lower than that of Figure \ref{fig:case1_a}. This means that the document
of Figure \ref{fig:case1_a} will be ranked higher
for the example query than the document on the right.
Intuitively speaking, the closer the query terms
are to the clusters of the constituent word vectors of a document, the higher is the similarity of the document with the query.

\begin{figure}[t]
    \subfloat[]{\label{fig:case2_a}
    \includegraphics[width=0.49\columnwidth]{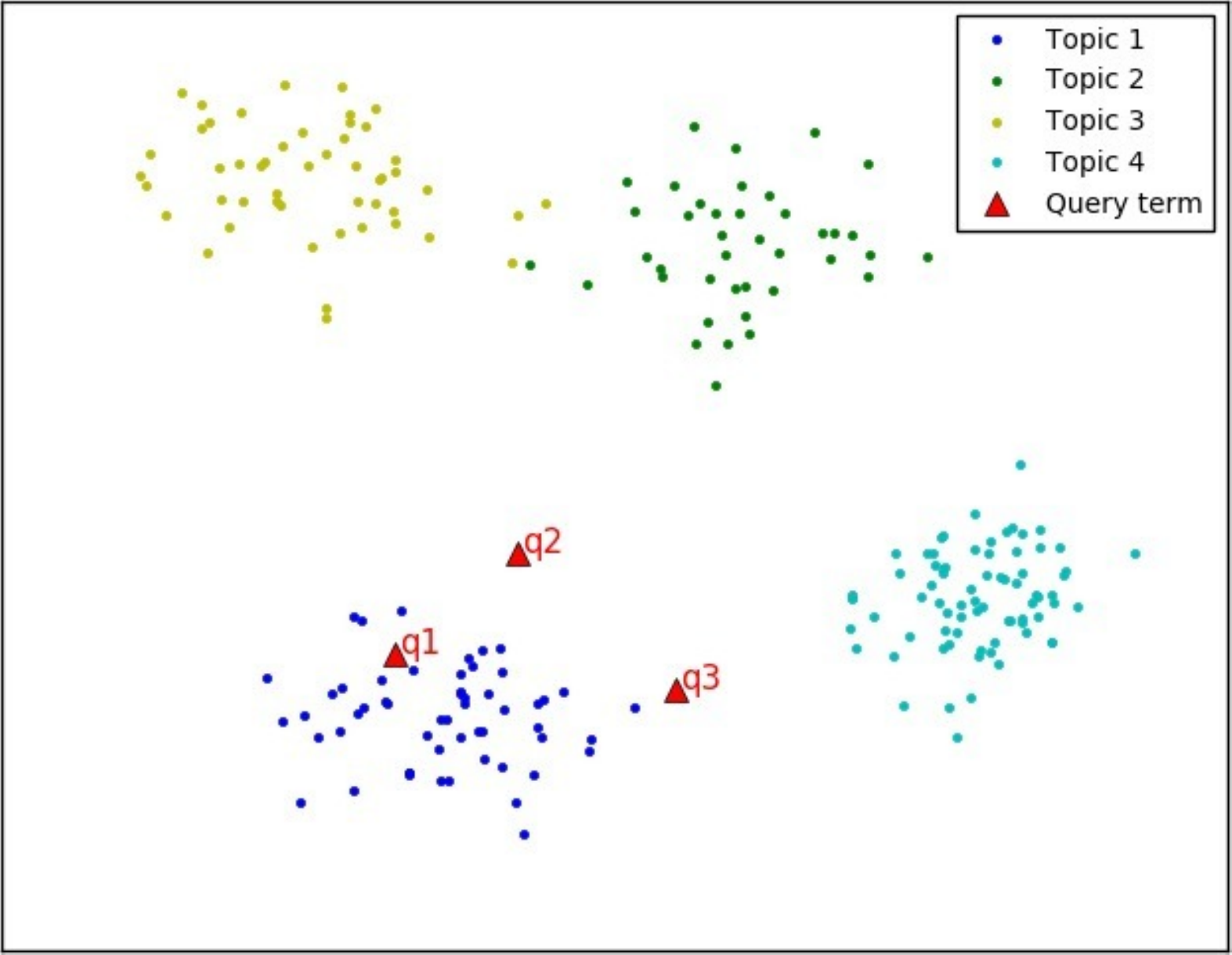}}
    \subfloat[] {\label{fig:case2_b}
    \includegraphics[width=0.49\columnwidth]{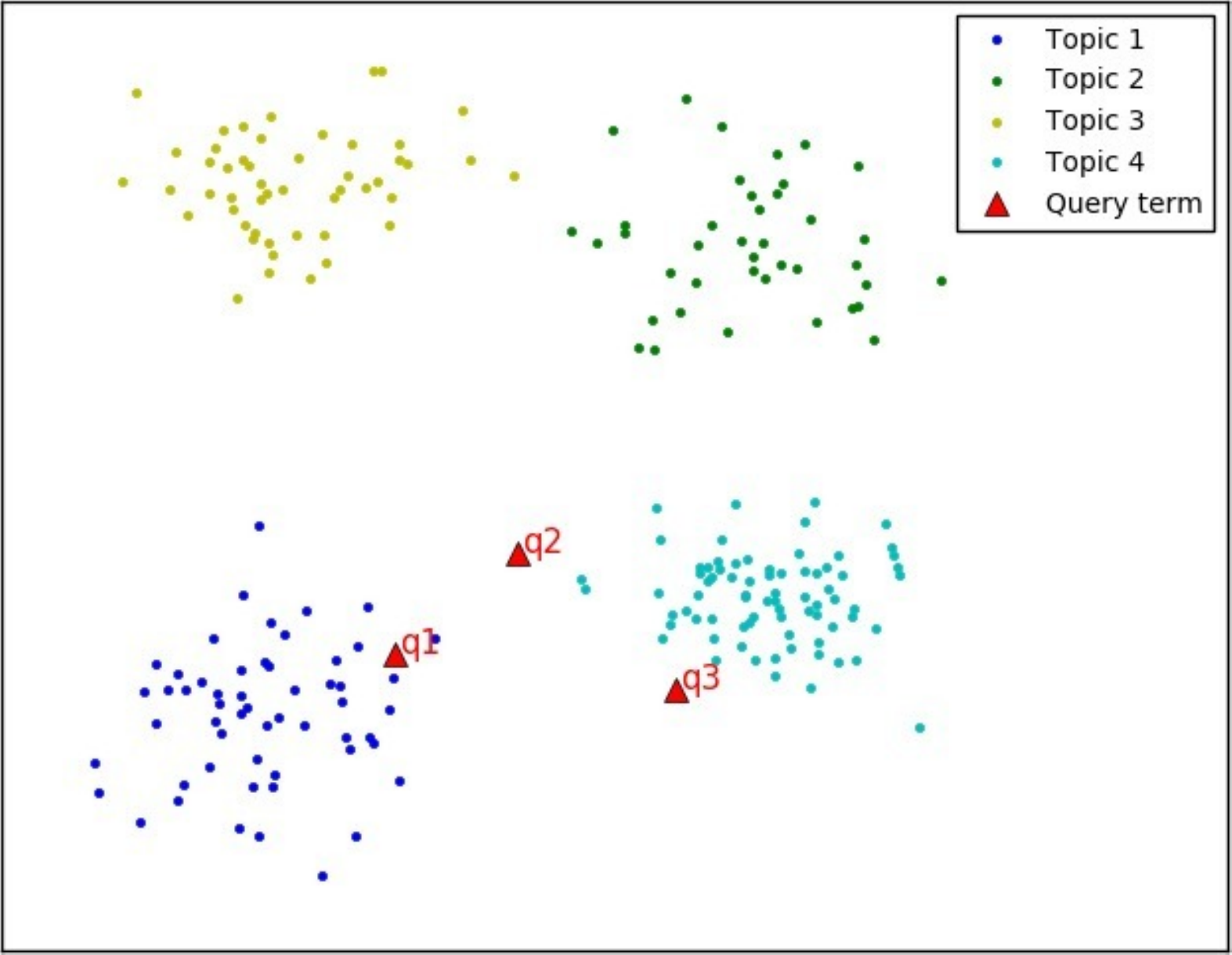}}
    \caption{Two example scenarios where documents are multi-topical, i.e. K-means clustering shows $4$ distinct clusters. \label{fig:case2}}
\end{figure}

Figure \ref{fig:case2} demonstrates two example cases where
$4$ distinct clusters of terms (topics) can be seen. 
The query likelihood for the document
in Figure \ref{fig:case2_a} is lower than that of the document
in Figure \ref{fig:case2_b}. This is because the observed query terms
in Figure \ref{fig:case2_b} are close to the cluster centres on the bottom-left and bottom-right, whereas for the example scenario in
Figure \ref{fig:case2_a}, the query point $q_2$ is relatively
far away from all the centroids. This has the implication that
the information need aspect of the query term $q_2$ is not
adequately expressed in the contents of the document in Figure \ref{fig:case2_a}.



\subsection{Combining with Text Likelihood}

Equation \ref{eq:avgsim} gives a new way to compute query likelihoods in comparison to the standard method of computing $P(d|q)$ using the BoW representation model using the standard language modeling (LM) \cite{hiemstra00,PonteThesis}.
In LM, the prior probability of generating a query $q$ from
a document $d$ is given by a Multinomial sampling probability
of obtaining a term $q_i$ from $d$ with prior belief $\lambda$ (a parameter for LM), coupled with a probability of sampling the term $q_i$ from the collection with prior belief $(1-\lambda)$.
Let this probability be represented by $P_{LM}(d|q)$
as shown in Equation \ref{eq:lm}.
\begin{equation}
P_{LM}(d|q) = \prod_{i}\lambda P_{MLE}(q_i|d) + (1-\lambda)P_{coll}(q_i) \label{eq:lm}
\end{equation}
In order to combine the standard LM query likelihood probability $P_{LM}(d|q)$ of Equation \ref{eq:lm} with the query likelihood estimated using the mixture model density estimate of the
word vectors, as shown in Equation \ref{eq:avgsim}, 
we introduce another indicator binary random variable to denote
the individual believes of the text based similarity and
the word vector based similarity. If the probability of
this indicator variable is denoted by $\alpha$, we obtain
the combined query likelihood as a mixture model
of the two respective query likelihoods, one for the text and
the other for the embedded word vectors.
\begin{equation}
P(d|q) = \alpha P_{LM}(d|q) + (1-\alpha)P_{WVEC}(d|q) \label{eq:simcomb}
\end{equation}

\section{Index Construction Details}  
\label{sec:indexing}



In our approach, as discussed in Section \ref{sec:representation}, the embedded representation of the
constituent terms of a document are used for estimating the similarity between a document and a query in combination with the BoW based LM similarity.
In this section, we discuss how can the constituent word vector representations be stored in the index so that the cluster centres for each document can be computed efficiently during retrieval time in order to compute its similarity with the
query.


\begin{table*}[t]
\centering
\begin{tabular}{@{}l@{~~}l@{~~}c@{~~}c@{~~}c@{~~}l@{~~}c@{~~}c@{~~}c@{~~}c@{~~}c@{}}
\toprule
Document & Document & \#Docs & Vocab & Query & Query Set & Query Ids & Avg qry & Avg \# & Dev & Test \\
Collection & Type & & Size & Fields & & & length & rel docs & Set & Set \\
\midrule
\multirow{3}{*}{TREC} & \multirow{4}{*}{News} & \multirow{4}{*}{528,155} & \multirow{4}{*}{242,036} & \multirow{4}{*}{Title} & TREC 6 ad-hoc & 301-350 & 2.48 & 92.2 & $\checkmark$ & \\ 
& & & & & TREC 7 ad-hoc & 351-400 & 2.42 & 93.4 & & $\checkmark$ \\ 
\multirow{1}{*}{Disks 4, 5} & & & & & TREC 8 ad-hoc & 401-450 & 2.38 & 94.5 & & $\checkmark$ \\ 
& & & & & TREC Robust & 601-700 & 2.88 & 37.2 & & $\checkmark$ \\
\bottomrule
\end{tabular}
\caption{\label{tab:trec}Dataset Overview}
\end{table*}

\begin{table*}[t]
\centering
\begin{tabular}{@{}l@{~~}l@{~~}c@{~~}c@{~~}l@{~~}l@{~~}l@{~~}l@{~~} l@{~~}l@{~~}l@{~~}l@{}}
\toprule
Dataset & Method & \multicolumn{2}{l}{Parameters} & \multicolumn{4}{c}{Initial Retrieval} & \multicolumn{4}{c}{RLM Feedback} \\

\cmidrule(r){3-4}
\cmidrule(r){5-8}
\cmidrule(r){9-12}

 & & \#clusters & $\alpha$ & MAP & GMAP & Recall & P@5 &
 MAP & GMAP & Recall & P@5 \\
\midrule

\multirow{4}{*}{TREC-6} & LM & n/a & n/a & \textbf{0.2363} & 0.0914 & \textbf{0.5100} & 0.4040  & \textbf{0.2656} & \textbf{0.0937} & \textbf{0.5386} & 0.4160 \\ 
& LM+wvsim$\mathrm{_{one\_cluster}}$ & 1 & 0.4 & 0.2355 & \textbf{0.0918} & 0.5058 & 0.3920 & 0.2538 & 0.0893 & 0.5279 & 0.4200\\
& LM+wvsim$\mathrm{_{no\_cluster}}$ & n/a & 0.4 & 0.2259 & 0.0827 & 0.5000 & 0.3600 & 0.2579 & 0.0894 & 0.5356 & 0.4440\\
& LM+wvsim$\mathrm{_{kmeans}}$ & 100 & 0.4 & 0.2345 & 0.0906 & 0.5027 & \textbf{0.4040} & 0.2537 & 0.0900 & 0.5338 & \textbf{0.4480} \\

\midrule
\multirow{4}{*}{TREC-7} & LM & n/a & n/a & \textbf{0.1787} & 0.0831 & 0.4882 & \textbf{0.4120} & 0.2095 & 0.0944 & 0.5481 & 0.4080 \\
& LM+wvsim$\mathrm{_{one\_cluster}}$ & 1 & 0.4 & 0.1773 & 0.0851 & 0.4897 & 0.3960 & 0.2077 & 0.0931 & 0.5566 & 0.4120 \\
& LM+wvsim$\mathrm{_{no\_cluster}}$ & n/a & 0.4 & 0.1664 & 0.0803 & 0.4863 & 0.3640 & 0.2104 & 0.0911 & 0.5549 & 0.4120 \\
& LM+wvsim$\mathrm{_{kmeans}}$ & 100 & 0.4 & 0.1756 & \textbf{0.0874} & \textbf{0.4916} & 0.3840 & \textbf{0.2125} & \textbf{0.0976} & \textbf{0.5616} & \textbf{0.4120} \\

\midrule

\multirow{4}{*}{TREC-8} & LM & n/a & n/a & 0.2462 & 0.1384 & 0.5932 & 0.4560 & 0.2689 & 0.1512 & 0.6436 & 0.4640 \\
& LM+wvsim$\mathrm{_{one\_cluster}}$ & 1 & 0.4 & 0.2541$^{\dagger}$ & 0.1465 & 0.6017 & 0.4440 & \textbf{0.2790} & \textbf{0.1588} & \textbf{0.6618} & \textbf{0.4920} \\
& LM+wvsim$\mathrm{_{no\_cluster}}$ & n/a & 0.4 & 0.2473 & 0.1396 & 0.5994 & 0.4520 & 0.2761 & 0.1576 & 0.6594 & 0.4880 \\
& LM+wvsim$\mathrm{_{kmeans}}$ & 100 & 0.4 & \textbf{0.2558}$^{\dagger}$ & \textbf{0.1468} & \textbf{0.6017} & \textbf{0.4720} & 0.2741 & 0.1588 & 0.6603 & 0.4880 \\
\midrule

\multirow{4}{*}{Robust} & LM & n/a & n/a & 0.2698 & 0.1724 & 0.7935 & 0.4485        & \textbf{0.3255} & \textbf{0.2080} & \textbf{0.8537} & \textbf{0.4990} \\
& LM+wvsim$\mathrm{_{one\_cluster}}$ & 1 & 0.4 & 0.2690 & 0.1701 & 0.7905 & 0.4465 & 0.3172 & 0.2023 & 0.8440 & 0.4808 \\
& LM+wvsim$\mathrm{_{no\_cluster}}$ & n/a & 0.4 & 0.2642 & 0.1646 & 0.7900 & 0.4485 & 0.3184 & 0.2012 & 0.8440 & 0.4808 \\
& LM+wvsim$\mathrm{_{kmeans}}$ & 100 & 0.4 & \textbf{0.2804}$^{\dagger}$ & \textbf{0.1819} & \textbf{0.8010} & \textbf{0.4687} & 0.3160 & 0.2002 & 0.8456 & 0.4848 \\

\bottomrule

\end{tabular}
\caption{Comparative results of set-based word vector similarities with different settings.
Parameters $K$ and $\alpha$ respectively denote the number of clusters and the relative weight of the text based query likelihood. $\dagger$ denotes significance with respect to text only similarity (LM).
\label{tab:res}
}
\end{table*}

\subsection{Pre-Clustering the Collection Vocabulary}

Word embedding techniques ensure that semantically related words are embedded in close proximity. This means that
the K-means clustering algorithm executed once for the whole vocabulary of the words in a collection is able to
cluster the words into distinct semantic classes. Each semantic class represents a global topic (cluster id of a term) of the whole collection.

In order to obtain per-document topics from these global topic classes, while indexing each document,
we perform a table look-up to retrieve the cluster id
of each constituent term. Note that since the cluster id, $c(w)$
of each term $w$ is an integer in $[1, K]$, the maximum number of unique cluster ids a document can have is $K$. In practice, the number of unique cluster ids of the constituent word vectors of a document will be much smaller than $K$, typically in the range of $5$ to $20$.

The words in a document are then grouped by their cluster ids,
and for each cluster id, we compute the cluster's centroid
by averaging the word vectors for that group. Formally speaking, from the BoW representation of a document $W_d = \{w_i\}_{i=1}^{d}$, we obtain a more compact representation
of $d$ by averaging the word vectors of each cluster group $C_k$ as shown in Equation \ref{eq:compactd}.
\begin{equation}
\mu_k = \frac{1}{|C_k|} \sum_{x \in C_k}x, C_k=\{x_i:c(w_i)=k\},i=1,\ldots,|d|   \label{eq:compactd}
\end{equation}
With Equation \ref{eq:compactd}, a document is represented
as a variable length vector of cluster centroids. The number of clusters in a document depends on its topical focus.
The index stores this additional information about the cluster centroids. During retrieval time, we compute the average
similarity between the query points and the stored centroid
vectors of a document.

\section{Evaluation} 
\label{sec:eval}

In this section, we report 
our experimental
investigation of the proposed method.
The objective of our experiments is to investigate whether incorporating semantic information (relative semantic similarities between the terms), when used in combination
with the standard text based similarity can further improve retrieval effectiveness.

\subsection{Experimental Setup}
\label{ss:setup}

Our experiments were conducted on the TREC ad-hoc test collections. The topic sets used for our experiments constitute the TREC 6, 7, 8 and Robust ad-hoc task topics.
The characteristics of the document and the query sets are
outlined in Table \ref{tab:trec}.

As our baseline retrieval model for initial retrieval, we used the standard LM with Jelinek Mercer smoothing \cite{hiemstra00,Zhai2004}, which is distributed 
as a part of Lucene\footnote{\url{https://lucene.apache.org/core/}}, an open source IR framework.
The proposed methods are also implemented as a part of Lucene\footnote{Available at \url{https://github.com/gdebasis/txtvecir}}.
The smoothing parameter for LM $\lambda$ is set to the optimal value of $0.4$ after varying it in the range $[0.1, 0.9]$.
Word vectors were embedded with the help of the \emph{word2vec} C tool \cite{Mikolov13}, trained on a pre-processed version of the TREC collection, i.e. stop word removed and stemmed (with Porter stemmer). 
The word embeddings on the respective collections were obtained by embedding them in a $200$ dimensional space using continuous bag-of-words with negative sampling, as per the parameter settings prescribed in \cite{Mikolov13}.


All the parameters for our experiments are trained on the
TREC 6 topic set and are tested with the same parameter
settings on the rest of the topic sets.
Two parameters, crucial to our method, are the number of
clusters, $K$, used for clustering the vocabulary of words,
these clusters being used subsequently to obtain the
document representations. After performing initial experiments
with this parameter by varying it in the range of $10$ to $300$, we noted that the best results are obtained with $K$ set to $100$, which is why we report our results with this particular setting of the $K$ parameter.

To see the usefulness of the coarse-grained clustered
representation of the word vectors for computing the
set-based similarity between a document and a query, we
include two extreme cases in our experimental results, one
which considers the extreme fine-grained representation
of each individual word as its own cluster, and the other
which considers the extreme coarse-grained representation
of averaging every constituent word of a document into a
single centroid point. These two cases are denoted by
the names LM+wvsim$\mathrm{_{no\_cluster}}$
and LM+wvsim$\mathrm{_{one\_cluster}}$ respectively
in our experiment results.

Another parameter for our proposed method is the relative
belief, denoted by the parameter $\alpha$, that we put into the text based similarity while
combining it with the vector based similarity, as
shown in Equation \ref{eq:simcomb}. After a grid search
in $[0.1, 0.9]$, we found out that the optimal value
of this parameter is $0.4$, and therefore
we report the results with this particular setting of $\alpha$.

To see the benefits of pseudo-relevance feedback (PRF) on the initial retrieval results,
we conducted additional experiments by applying relevance model based feedback \cite{lavrenko_croft2001,rm3} on both the sets of initial retrieval results, i.e. the baseline LM and LM in combination with word vector based query likelihood. 

\subsection{Results} 
\label{ss:results}

In this section, we report the empirical performance of the proposed method on each dataset. The results of our experiments are reported in Table
\ref{tab:res}. We note that the results with $K=100$ are
the best for the TREC 8 and the TREC Robust topic sets.
In fact, these results are significantly better than the
results obtained with only text based similarity, which
in this case is the standard LM retrieval model.
TREC-8 and TREC Robust results show consistent improvements
in both recall and precision at top ranks.

We also note that on the TREC 8 and the Robust topic sets, the clustered representation of the documents produces much better results in comparison to the method of considering a very fine-grained representation of a document by using each constituent word vector for the similarity computation with the query. Somewhat surprisingly, the results with $K=1$, i.e.,
representing each document by a single point (the average
of each word in the document) is quite close in performance
to $K=100$, which makes us believe that we need to use
much higher values of $K$ for clustering the whole vocabulary set since it is likely that $100$ concepts or semantic classes
is too small to capture the wide diversity in the vocabulary
of the TREC collection.

Another observation is that the results with $K=100$ are not the best for
the TREC 6 and the TREC 7 topic sets. The results with
$K=100$ show marginal improvements. Surprisingly,
the single-point representation of the documents produces better results than with $K=100$.

From Table \ref{tab:res}, we observe that the initial retrieval results
obtained with word vector based likelihood are in general
improved by application of RLM based PRF, e.g. the TREC 7 and the TREC 8 results.
However, for TREC 6 and Robust, the best PRF results are obtained with the LM based
initial retrieval.

\begin{figure}[t]
  \centering
  \includegraphics[width=.49\columnwidth]{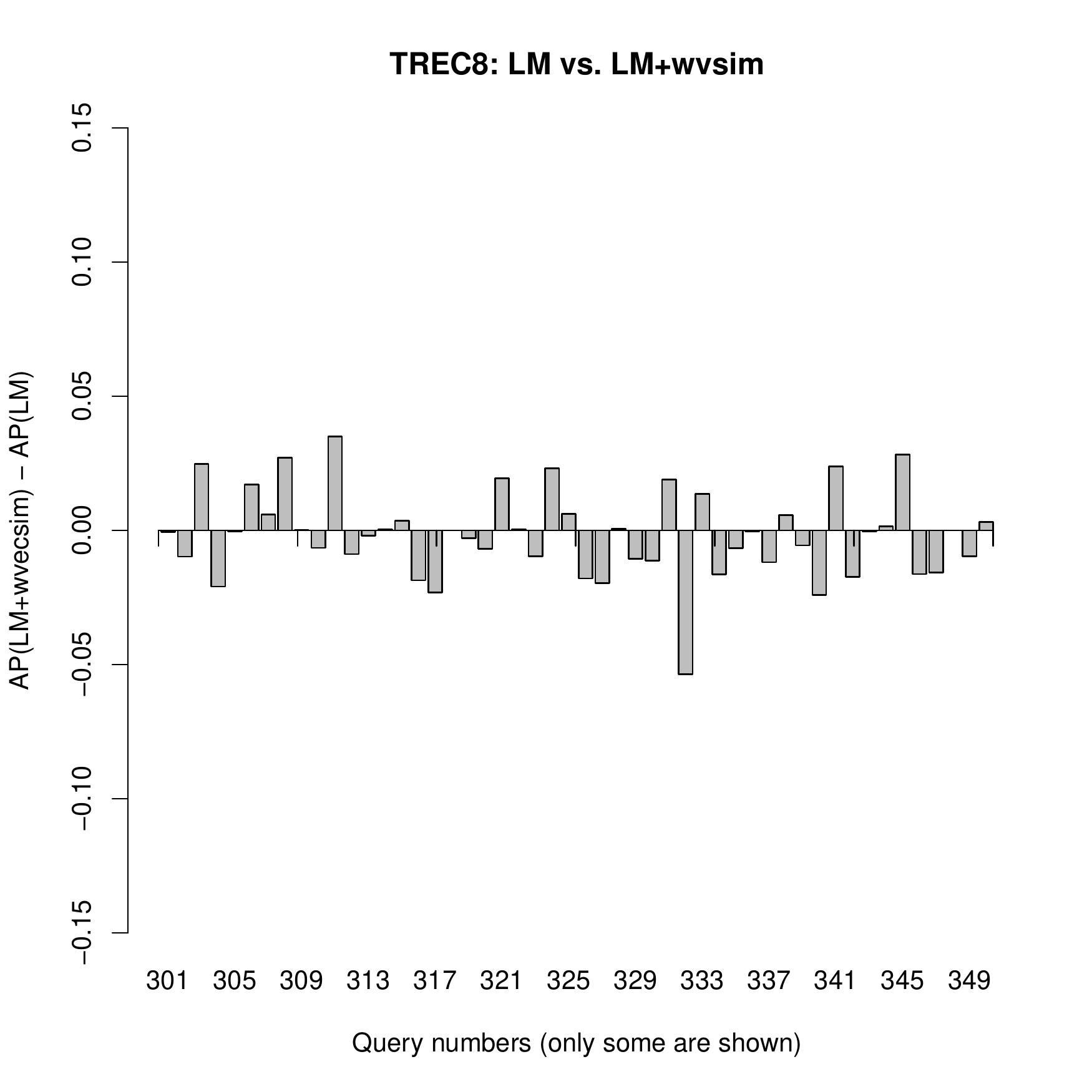}
  \includegraphics[width=.49\columnwidth]{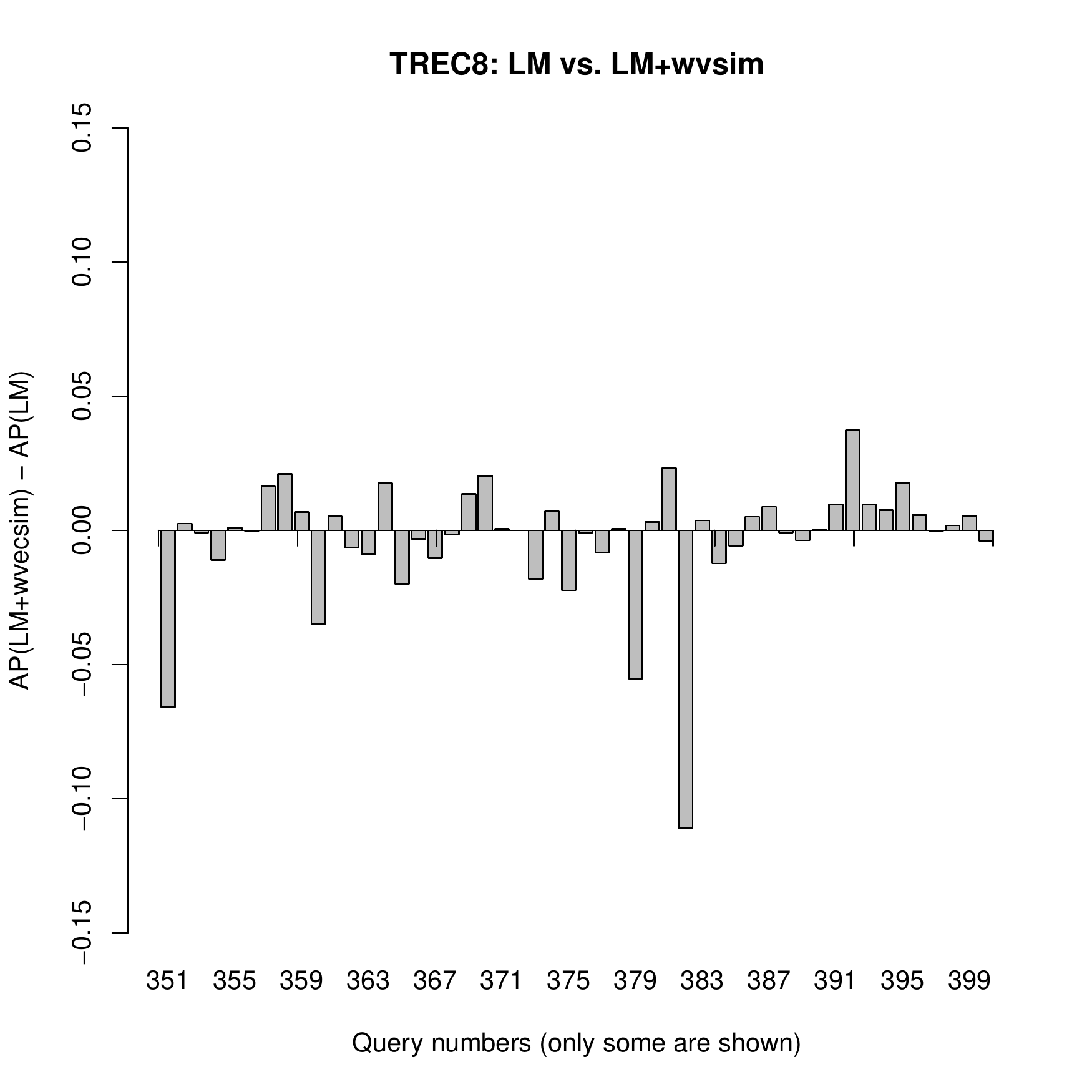} \\
  \includegraphics[width=.49\columnwidth]{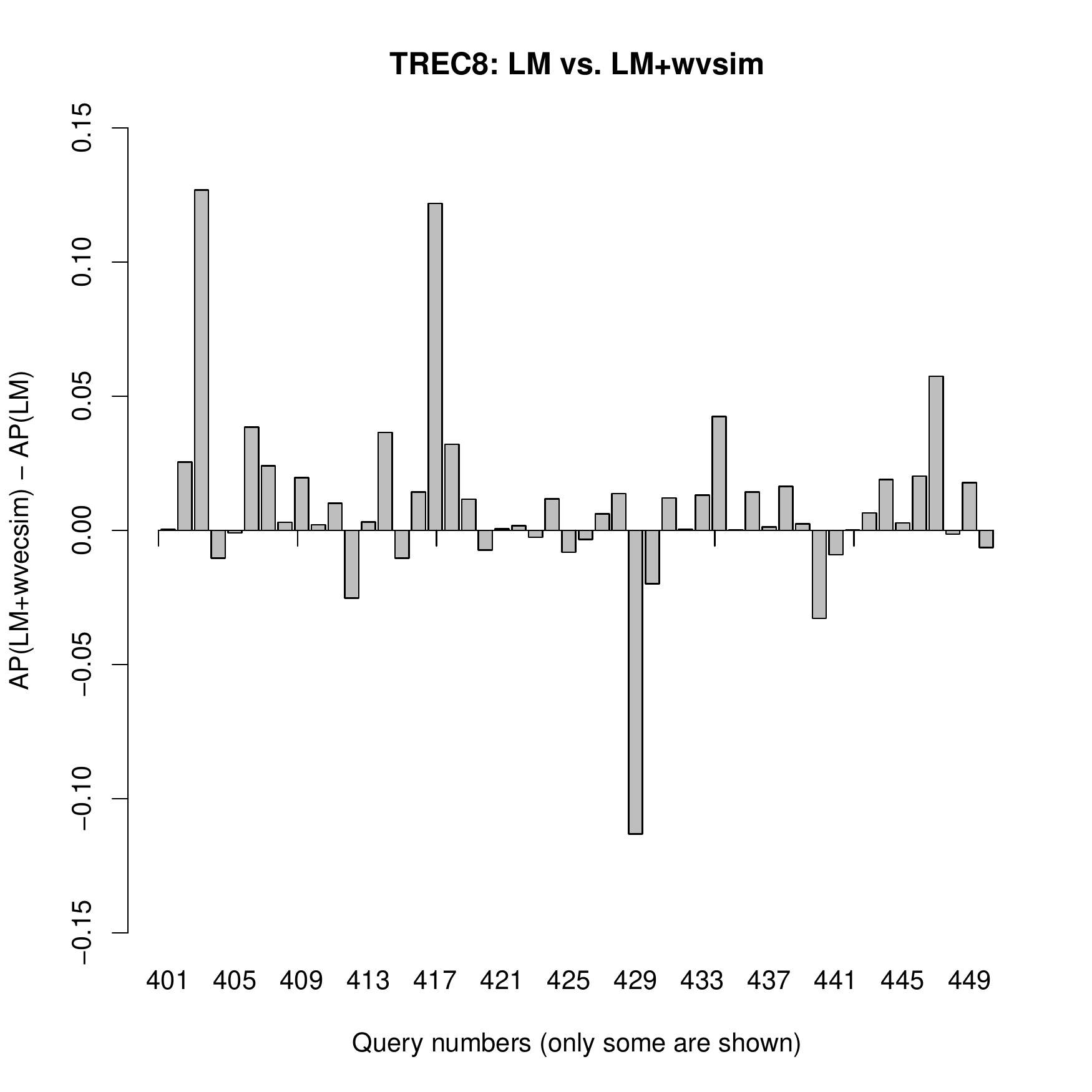}
  \includegraphics[width=.49\columnwidth]{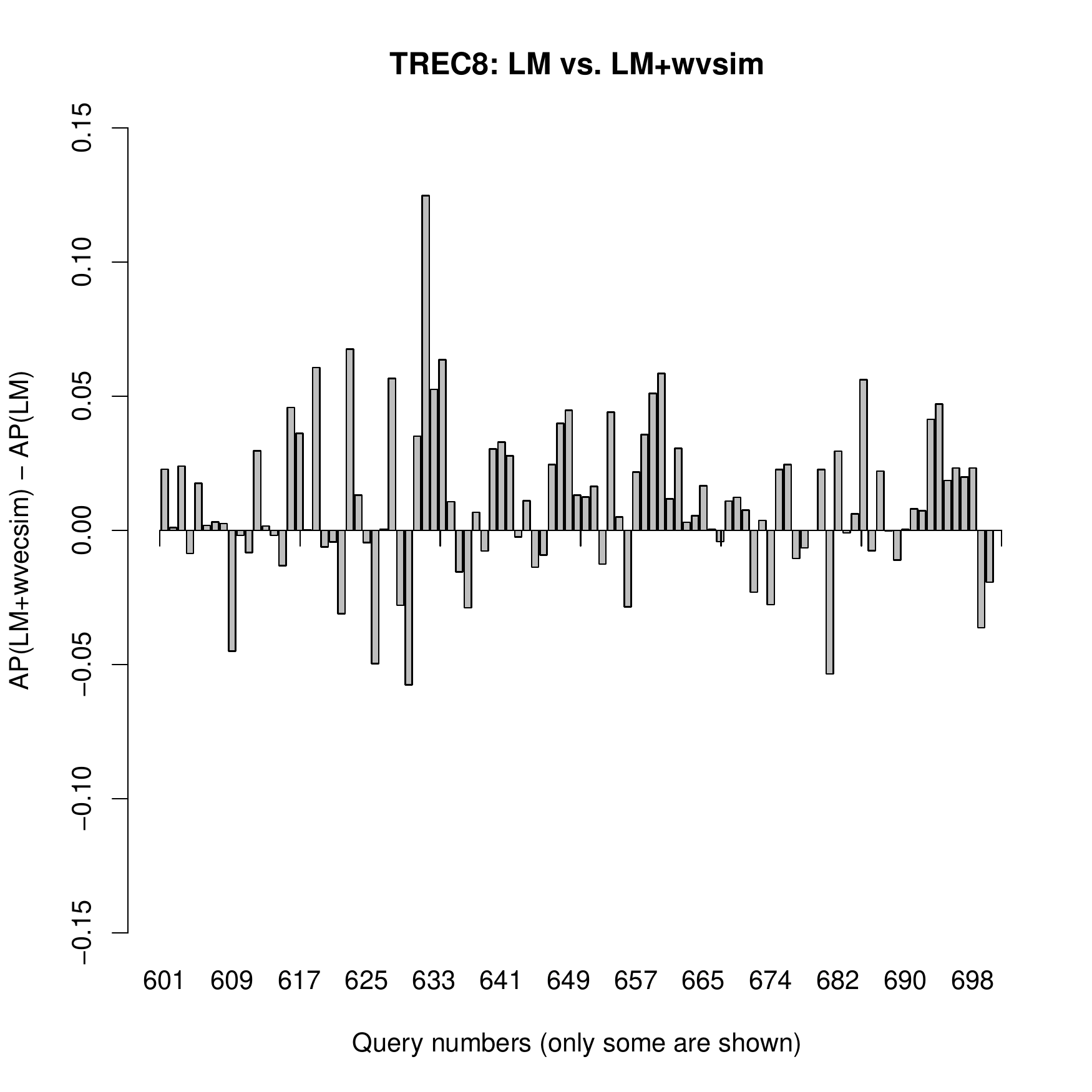}
  \caption{Per-query differences in AP values between LM and the combination approach of LM with word vector based likelihood ($\#clusters=100$)}
  \label{fig:qq}
\end{figure}

Figure \ref{fig:qq} investigates the effect of word vector based query likelihood on individual queries. Each vertical bar in the plots of Figure \ref{fig:qq}
denotes the difference between the average precision (AP) values of
the retrieval results obtained with the combination of LM with the word vector based similarity and LM alone. Ideally, we would want the vertical bars to be higher over X axis. 
From Figure \ref{fig:qq} it appears that the combination turns out to be beneficial for a number of queries while hurting
the performance of others. It may be argued that the results can further be improved with selective application of the word vector based query likelihood. In order to see how much benefit we can get in the ideal situation, we assume the existence of an \emph{oracle} which always correctly 'selects' the best approach, i.e. chooses the best result given two approaches, one the baseline LM and the other with the word vector based likelihood in combination with it. Table \ref{tab:oracle} shows the best results that can, in
principle, be achieved with using the word vector based similarities, on each
topic set.

\begin{table}[t]
\small
\centering
\begin{tabular}{lcc}
\toprule
Dataset & \multicolumn{2}{c}{MAP} \\
& wvsim & wvsim$^*$ \\
\midrule
TREC6 &  0.2345 & 0.2414 \\
TREC 7 & 0.1756 & 0.1837  \\
TREC 8 & 0.2558 & 0.2607 \\
TREC Robust & 0.2804 & 0.2833 \\
\bottomrule 
\end{tabular}
\caption{The best performance achievable, in principle, with the word vector based IR.
`wvsim' denotes the results obtained with $K=100$ and `wvsim$^*$' denotes
the Oracle approach.}
\label{tab:oracle}
\end{table}

\section{Conclusions and Future work} 
\label{sec:concl}

In this paper, we have proposed a novel approach to represent queries and documents as sets of embedded word vectors.
A document is represented as a probability density function
of a mixture of Gaussians. The posterior query likelihood
is estimated by the average distance of the query points from
the centroids of these Gaussians.
This word vector based query likelihood is then combined with
the standard LM based query likelihood for document ranking.
Experiments on standard text collections showed that the combined similarity measure almost always outperforms (often significantly) the LM (text based) similarity measure.

As a possible future work, we would like to apply our proposed representation of queries and documents for relevance feedback and query expansion. We would also like to explore various notions of distance measures (e.g. \cite{weinberger_distance}).
It would also be interesting to see the performance of this set-based representation obtained with a multi-sense word embedding method, such as \cite{neelakantan_efficient}.

\section{Acknowledgements}

This research is supported by Science Foundation Ireland (SFI) as a part of the ADAPT Centre at DCU (Grant No: 13/RC/2106) and by a grant under the SFI ISCA India consortium.

\bibliographystyle{abbrv}
{
\bibliography{wvs}
}
\end{document}